\documentclass[twocolumn,showpacs,prb]{revtex4}

\usepackage{graphicx}
\usepackage{dcolumn}
\usepackage{bm}

\begin{document}


\title{Superfluidity of $^4$He Confined in Nano-Porous Media}

\author{Keiya Shirahama}
 \email{keiya@phys.keio.ac.jp}
\author{Keiichi Yamamoto}
\author{Yoshiyuki Shibayama}

\affiliation{%
Department of Physics, Keio University, Yokohama 223-8522, Japan
}%

\date{\today}

\begin{abstract}
We have examined superfluid properties of $^4$He confined to a nano-porous Gelsil glass 
that has nanopores 2.5 nm in diameter.
The pressure-temperature phase diagram was determined by torsional oscillator, heat capacity and pressure studies. 
The superfluid transition temperature $T_{\mathrm c}$ approaches zero at 3.4 MPa, indicating a novel "quantum" superfluid transition.
By heat capacity measurements, the nonsuperfluid phase adjacent to the superfluid and solid phases is identified to be
a nanometer-scale, localized Bose condensation state, in which global phase coherence is destroyed.
At high pressures, the superfluid density has a $T$-linear term, and $T_{\mathrm c}$ is proportional to the zero-temperature 
superfluid density. These results strongly suggest that
phase fluctuations in the superfluid order parameter play a dominant role on the phase diagram and superfluid properties.  
\end{abstract}

\maketitle

\section{Introduction}
\begin{figure}[t]
\centering
 \includegraphics[height=80mm]{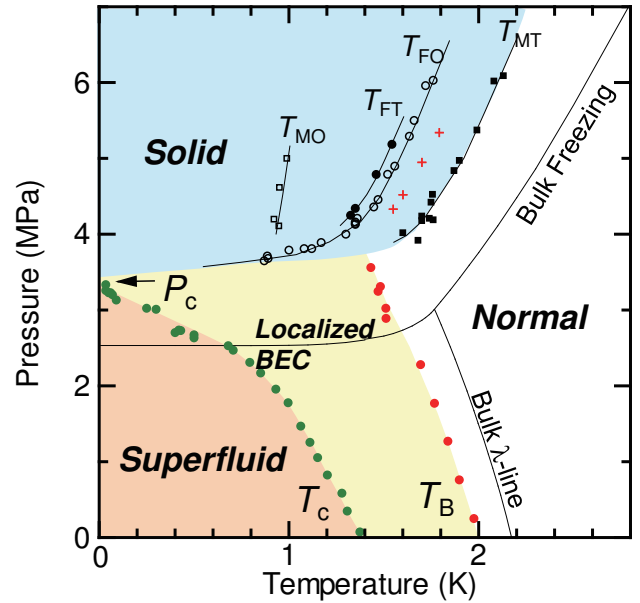}
  \caption{\label{PhaseDiagram}
$P-T$ phase diagram. 
The superfluid transition temperatures $T_{\mathrm c}$ (green dots) are obtained by torsional oscillator\cite{YamamotoPRL04} studies,
and the localized BEC temperatures $T_{\mathrm B}$ (red dots) and the melting points (red crosses) are by heat capacity\cite{YamamotoJLTP07,YamamotoPRL07}. 
Pressure and thermal response measurements\cite{YamamotoAIP05,YamamotoJPSJ07} give the melting and freezing lines: 
$T_{\mathrm {MO}}$; melting onset, $T_{\mathrm {MT}}$; melting termination, $T_{\mathrm {FO}}$; freezing onset, $T_{\mathrm {FT}}$; freezing termination,
respectively. Arrow indicates the critical pressure $P_{\mathrm c}$
at which $T_{\mathrm c}$ tends to 0 K.
}
\end{figure}

$^4$He confined or adsorbed in nanoporous media is an interesting model system of strongly correlated Bosons 
under external potential. 
Recently, we have investigated the superfluid and thermodynamic properties of $^4$He in 
nanoporous Gelsil glass, and have found that the strong confinement into the nanopores
causes a dramatic change in the phase diagram
\cite{YamamotoPRL04,YamamotoAIP05,YamamotoJPSJ07,YamamotoJLTP07,YamamotoPRL07,ShirahamaAIP05,ShirahamaTOP05,ShirahamaJLTP07}.
We show the obtained $P$-$T$ phase diagram in Fig. \ref{PhaseDiagram}.
With increasing pressure, the superfluid transition temperature $T_{\mathrm c}$ approaches zero at 3.4 MPa.
Measurements of isochoric pressure have suggested that
the freezing pressure is at or above 3.4 MPa\cite{YamamotoAIP05,YamamotoJPSJ07}.
These behaviors indicate a quantum phase transition (QPT) among superfluid, nonsuperfluid and solid phases induced by pressure as an
external parameter. 

The QPT behavior and the existence of a nonsuperfluid phase between the superfluid and solid phases
are in striking contrast to the case of bulk $^4$He. 
To investigate the nature of the nonsuperfluid phase and to understand the mechanism 
of QPT, we have made measurements of heat capacity and isochoric pressure\cite{YamamotoJPSJ07,YamamotoPRL07}.  
In this paper, we summarize the recent experimental results and propose an interpretation 
that will provide a novel perspective to the physics of $^4$He in porous media: 
The confinement of $^4$He to nanopores fluctuates the phase of superfluid order parameter,
and the phase fluctuations results in the localization of Bose-Einstein condensates and 
a quantum phase transition.

\section{Results: Phase Diagram and Superfluid Properties}
\subsection{Phase Diagram}
Here we summarize the results of the measurements of pressure and heat capacity,
and describe in more detail about the torsional oscillator studies, focusing on the temperature dependence of 
the superfluid fraction. The details of the results have been described elsewhere\cite{YamamotoPRL04,YamamotoAIP05,YamamotoJPSJ07,YamamotoJLTP07,YamamotoPRL07,ShirahamaAIP05,ShirahamaTOP05,ShirahamaJLTP07}.

We have employed a porous Gelsil glass\cite{Gelsil},
which is manufactured by the sol-gel process.
Gelsil has three-dimensionally (3D) interconnected nanopores, similarly to Vycor.
The nominal pore diameter of our glass samples is 2.5 nm.   
Since various pore sizes are available, Gelsil has been recently used in helium studies.
The controllability and wide variety of the pore size were not available in Vycor.
It was first employed for $^4$He study by Miyamoto and Takano (MT)\cite{Miyamoto96}. 
They found that the superfluid transition in a 2.5-nm Gelsil sample was depressed to 0.9 K at ambient pressure.

We have constructed a heat capacity cell having a capacitance pressure gauge\cite{YamamotoJLTP07,YamamotoPRL07}. 
This cell enables us to measure the pressure and heat capacity for the same glass sample.
The sample cell contains four Gelsil disk samples (5.5 mm diameter, 2.3 mm thick) 
which are taken from the same batch as the one used in the torsional oscillator experiments.

In the pressure study\cite{YamamotoAIP05,YamamotoJPSJ07}, we measure pressure $P(T)$ along isochores. The rates of cooling and warming of the cell 
are also recorded simultaneously. 
Both data show some signatures that are related to freezing and melting of $^4$He in the nanopores.
The freezing and melting occur at different temperatures and in finite temperature ranges, unlike the first order transition of bulk $^4$He.
We have identified $P(T)$ at which $^4$He starts and terminates to freeze and thaw. The four data sets are plotted in Fig. \ref{PhaseDiagram}.
The reduction of freezing and melting temperatures observed above 3.7 MPa.
Below 3.4 MPa, no signatures indicating freezing and melting were observed. 
The liquid-solid boundary below 0.8 K should therefore be located between 3.7 and 3.4 MPa, 
meaning that the freezing line is nearly flat and the entropy difference between the solid and nonsuperfluid phases is small.
This fact strongly suggests that the nonsuperfluid phase is a sort of an ordered state.

To clarify the nature of the low-entropy nonsuperfluid state, we have conducted the heat capacity measurement\cite{YamamotoJLTP07,YamamotoPRL07}.
In Fig. \ref{CRhoS}(a) is shown the heat capacity data, 
in which $^4$He in the nanopores is liquid.
A broad, but substantial peak is found in each heat capacity. The peak temperature $T_{\mathrm B}$ 
indicated by arrows and the peak height decrease as $P$ increases. 
We plot $T_{\mathrm B}$ on the phase diagram of Fig. \ref{PhaseDiagram}.
Obviously, the "$T_{\mathrm B}$ line" is located about 0.2K below the $\lambda$ line, and is parallel to the $\lambda$ line.

The heat capacity peak is reminiscent of the superfluid size effect in $^4$He in various restricted geometries\cite{Gasparini}.
However, the system exhibits no superfluid transition at and just below $T_{\mathrm B}$.
This is clearly seen in Fig. \ref{CRhoS}(b),
the data of the frequency shift in the torsional oscillator measurement.
In Fig. \ref{CRhoS}(b), small upturns seen in both data around 2 K are due to the superfluid transition 
of the bulk liquid in the open space of the cell.
The large, abrupt increase at lower temperatures indicates the superfluid transition of $^4$He confined in the nanopores.
The superfluid transition temperature $T_{\mathrm c}$ is much lower than $T_{\mathrm B}$, 
and it decreases progressively with increasing pressure.
The remarkable difference in the behaviors of two characteristic temperatures is obviously seen in Fig. \ref{PhaseDiagram}.  

\subsection{Superfluid Properties}
Torsional oscillator technique\cite{TorOsc} is based on a simple relationship that the frequency shift $\Delta f$ is proportional to the 
superfluid density $\rho _{\mathrm s}$. Therefore, $\Delta f(T)$ should contain 
essential information for understanding the nature of superfluidity. 
In the next section we focus on the behavior of $\rho _{\mathrm s}(T)$. Here we mention 
some features in the $\Delta f(T)$ curves in 
two density regions: (1) adsorbed films to filled-pore states, (2) pressurized states 
at $0 < P < 3.4$ MPa.

\subsubsection{Film States}
The adsorbed film shows the superfluid response when the coverage exceeds the critical coverage
$n_{\mathrm c} = 19.9~\mu {\mathrm {mol/m^2}}$.
The superfluid transition temperature $T_{\mathrm c}$ increases almost linearly with the superfluid coverage $n-n_{\mathrm c}$,
and has a maximum at $n_{\mathrm {full}} = 33~\mu {\mathrm {mol/m^2}}$, at which the pore is filled with $^4$He.
It should be noted that the amount of the nonsuperfluid (i.e. "inert") layers adjacent to the pore walls are larger than 
that of superfluid liquid under ambient pressure. The "effective" pore diameter for the superfluid part is therefore reduced 
to about 1.5 nm. 

From the slope of $f(n)$ in the nonsuperfluid and superfluid states, 
we obtained the ratio of undetected superfluid mass to total superfluid mass, 
the so called $\chi$ factor, to be 0.1. 
This value is much smaller than 0.33 in the case of Vycor, 
but larger than the obtained value by MT for the similar Gelsil glass, 0.06. 
These results indicate that the nanopores in Gelsil are more tortuous than the pores in Vycor. 
\begin{figure}[t]
\centering
 \includegraphics[width=80mm]{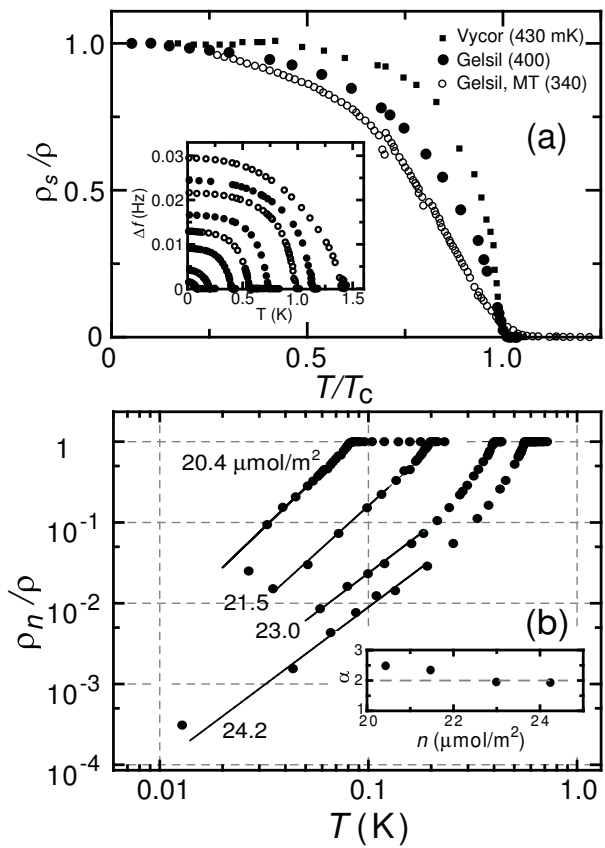}
  \caption{\label{FilmRhoSN}
Torsional oscillator results in the adsorbed film states. 
(a) Comparison of the normalized superfluid fraction ($\rho _{\mathrm s}/ \rho$ versus $T / T_{\mathrm c}$)
to the data of $^4$He in other 2.5-nm Gelsil\cite{Miyamoto96} and in Vycor\cite{Bishop81}. 
The adopted data have similar $T_{\mathrm c}$.
Inset: Frequency shift $\Delta f(T)$ for eight coverages. 
(b) Log-log plots of the normal fluid fraction $\rho _{\mathrm n}(T) / \rho$ for
four coverages. The solid lines are the best powerlaw fittings.
Inset: Powerlaw exponents $\alpha$ obtained in the fitting 
$\rho _{\mathrm n}(T) / \rho = a T^{\alpha}$ as a function of $n$.}
\end{figure}

In Fig. \ref{FilmRhoSN}, we show the typical superfluid frequency shift normalized by the shift at 0 K, 
$\Delta f(T)/\Delta f(0)$, which is equal to the superfluid fraction $\rho_{\mathrm s}(T)/\rho$, 
together with the similar data of $^4$He film in 2.5-nm Gelsil by MT and in Vycor.
The $\Delta f$ data set at various coverages are shown in the inset of Fig. \ref{FilmRhoSN}(a).
There exists substantial difference in the temperature dependence of $\rho_{\mathrm s}(T)/\rho$ among three experiments. 
Our Gelsil data lies between the Vycor data and Gelsil data by MT, and possess the features of 
these two systems.
In the Vycor case, $\rho_{\mathrm s}(T)$ is proportional to $T^2$ at low temperatures,
and show a bulk-like critical behavior near $T_{\mathrm c}$\cite{Bishop81}.
The $T^2$ behavior suggests that one-dimensional phonons are the dominant low-energy excitations. 
In the Gelsil experiment by MT\cite{Miyamoto96}, $\rho_{\mathrm s}(T)/\rho$ is also fitted to 
$T^2$ at low temperatures at $T < 0.4 - 0.8 T_{\mathrm c}$. 
As shown in Fig. \ref{FilmRhoSN}(b), also in our Gelsil the normal fluid density obeys 
approximately $T^2$ law at low temperature regions, as shown in the inset.
Near $T_{\mathrm c}$, our $\rho_{\mathrm s}(T)$ resembles the Vycor data, 
although the data are not enough to accurately determine the critical exponent.  

\subsubsection{Liquid under Pressure}
\begin{figure}[t]
\centering
 \includegraphics[width=70mm]{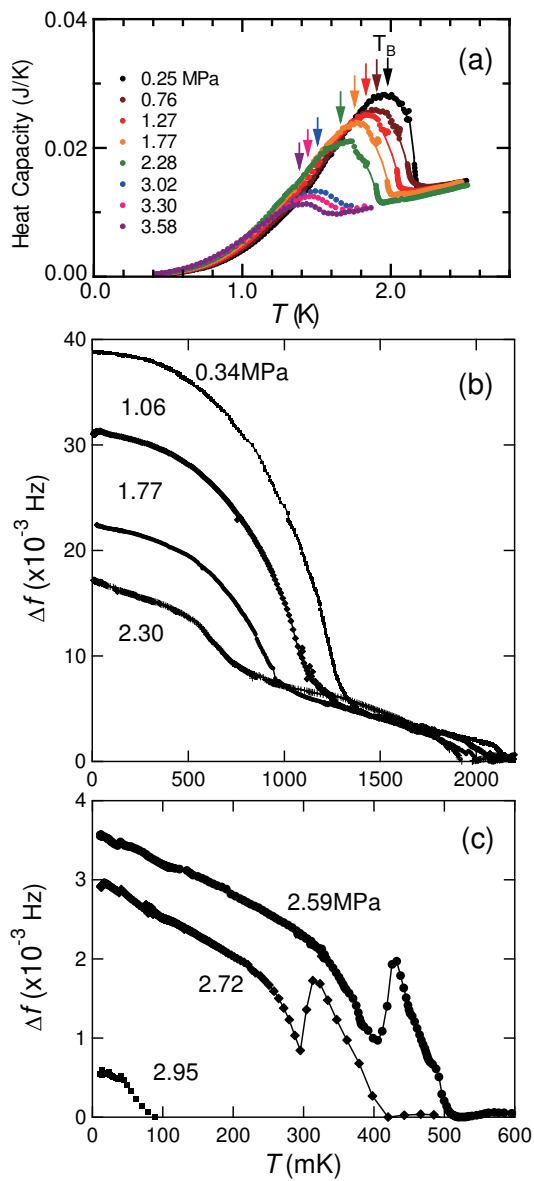}
  \caption{\label{CRhoS}
(a) Heat capacity of liquid $^4$He in the Gelsil nanopores for eight pressures.
Arrows indicate the peak temperatures $T_{\mathrm B}$ that are interpreted
as the LBEC formation temperatures. See recent publication\cite{YamamotoJLTP07,YamamotoPRL07} for
method of extraction of the heat capacity in the nanopores.
(b) Torsional oscillator frequency shifts $\Delta f(T)$ at pressures below the bulk melting pressure.
The shifts starting around 2 K are the contribution from bulk liquid in the sample cell. 
(c) $\Delta f(T)$ at pressures above the bulk melting pressure. The $n$-shaped
anomalies are anti-crossing resonances resulting from the coupling to superfluid fourth sound.}
\end{figure}

Next we mention the results obtained in the pressurized states, where liquid $^4$He fills the nanopores\cite{YamamotoPRL04}. 
In Fig. \ref{CRhoS}(b) and (c) are shown the $\Delta f(T)$ data. 
The data in Fig. \ref{CRhoS}(b) are obtained by subtraction of the empty cell background 
at bulk superfluid trantision temperatures. 

\begin{figure}[t]
\centering
 \includegraphics[width=85mm]{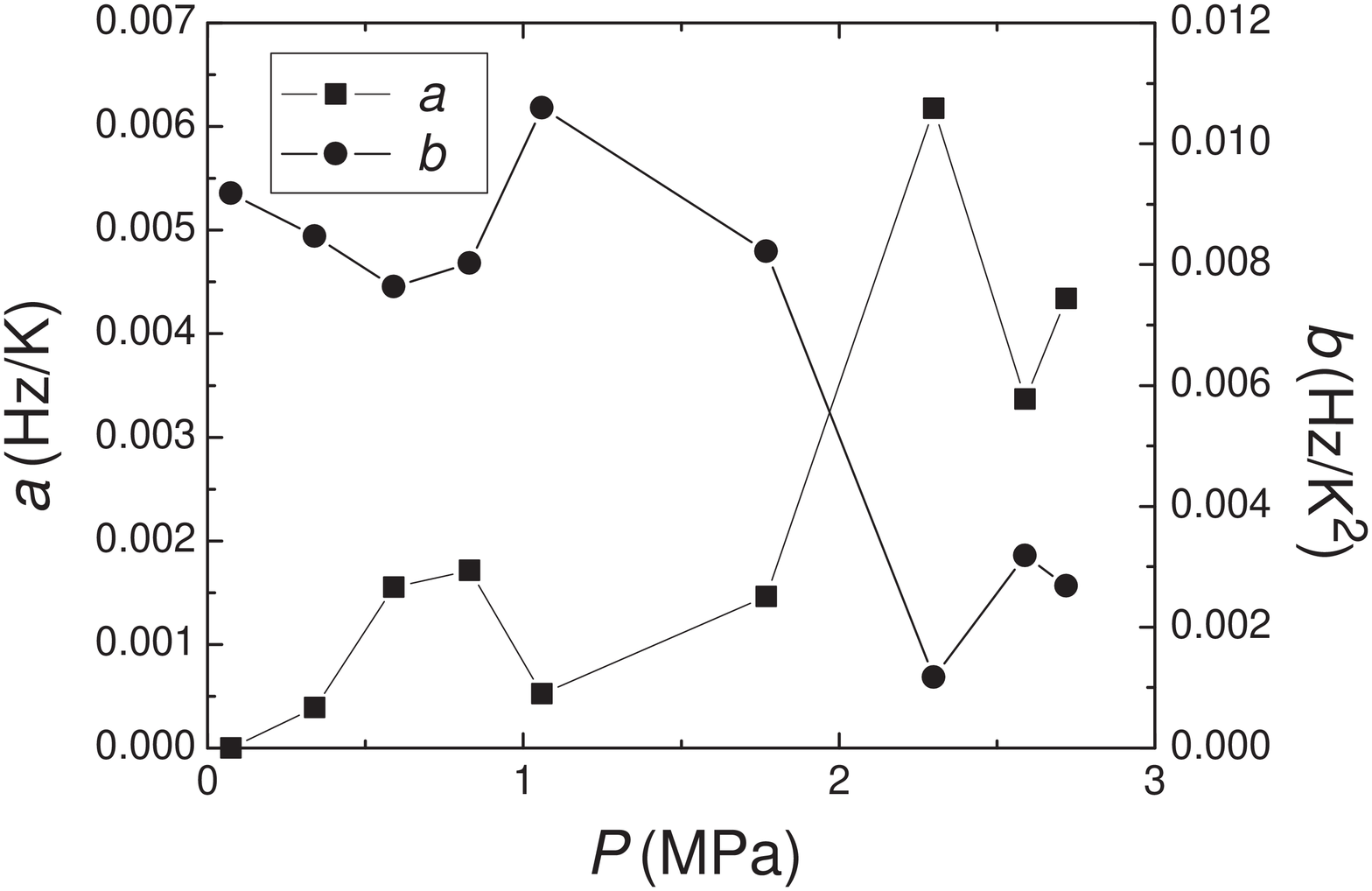}
  \caption{\label{Coefficient}
The coefficients $a$ and $b$ obtained from the linear-parabola fitting $\Delta f(0) - \Delta f(T) = aT + bT^2$ 
for the data taken at nine pressures.}
\end{figure}

We have found that also in pressurized liquid $\Delta f(T)$ obeys powerlaw at low temperatures.
At $P < 1.7$ MPa the normal fluid fraction $\rho _{\mathrm n}/\rho$ is best fitted by
$\rho _{\mathrm n}/\rho \propto T^{\beta}$ with the exponent $\beta$ ranging from 2.3 to 2.5.
At higher pressures, a $T$-linear behavior emerges.
In order to see the crossover from the nearly parabolic to linear temperature dependence,
we have fitted the normal fraction to the sum of $T$-linear and square terms,
$\rho _{\mathrm n}/\rho = aT + bT^2$. 
The obtained coefficients $a$ and $b$ are plotted in Fig. \ref{Coefficient}.
Obviously, the $T$-linear term dominates $\rho _{\mathrm n}/\rho$ above 2.3 MPa.
We will discuss the origin of the $T$-linear term in the next section.
\section{Discussion: Localized BEC and Phase Fluctuation}
\subsection{The Localized BEC}
We have proposed in the previous publications that the QPT behavior, i.e. the anomalous reduction in superfluid $T_{\mathrm c}$,
results from the localization of Bose-Einstein condensates in the nonsuperfluid state\cite{ShirahamaAIP05,ShirahamaTOP05,ShirahamaJLTP07}.
We believe that this conjecture is now proven by the heat capacity measurement\cite{YamamotoJLTP07,YamamotoPRL07}.

\begin{figure}[t]
 \centering
 \includegraphics[height=50mm]{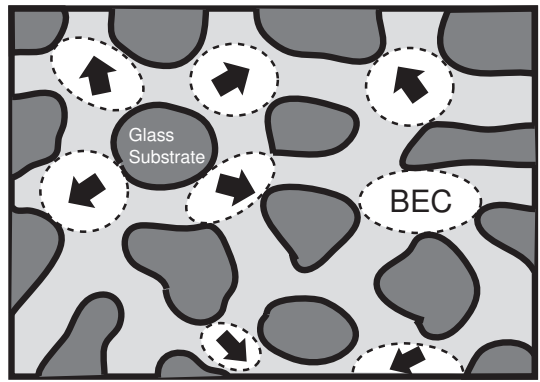}
  \caption{\label{LBEC}A cartoon showing the formation of localized BECs (LBEC) (shown as white) in a porous glass substrate (dark grey). 
  $^4$He atoms form many small BEC's at the wider regions
  (especially intersections of the pores), where the atoms can exchange actively. 
  The phase of each BEC is illustrated by thick arrows. 
  Since no phase coherence exist among the LBECs due to 
  the hard core of $^4$He suppressing the spatial exchanges at the narrower regions (light grey),
  the whole system exhibits no superfluidity on macroscopic length scale.
  As temperature is lowered, thermal phase fluctuations are diminished, 
  and the system should undergo a macroscopic superfluid transition at some temperature $T_{\mathrm c}$.}
\end{figure}

The idea of the localized BEC (LBEC) is shown in a cartoon of Fig. \ref{LBEC}:
When liquid $^4$He is confined in the nanopores, 
the BEC transition temperature should be reduced below bulk $T_{\lambda}$ due to the size effect. 
 Around a certain temperature below $T_{\lambda}$, many BECs 
 grow from large pores or intersections of pores, in which $^4$He atoms can exchange frequently their positions. 
 The size of the BECs is roughly limited to the pore size. 
 The atom exchange between the BECs via the narrow regions of the pores are interrupted,
 because $^4$He atom has a hard core. For the movement of one $^4$He atom, the surrounding $^4$He atoms act as a potential.
 The lack of the atom exchanges causes fluctuations in phase of the superfluid order parameter.  
 Therefore, no phase coherence exists among the BECs, and the whole system has also no global phase coherence
 and does not exhibit superfluidity that can be detected by macroscopic and dynamical measurements such as torsional oscillator.
  As the temperature is further decreased, the phase coherence between the localized BECs grows,   
 and macroscopic superfluidity, which is detected by torsional oscillator technique, is realized when most of the BECs coalesce.

The heat capacity peak provides a definite evidence for the formation of LBECs at $T_{\mathrm B}$.
Broad peak structure in heat capacity is a common feature of $^4$He in restricted geometries, and was recognized as 
a manifestation of superfluidity and BEC. The temperature dependence of the heat capacity (the shape of the peak)
of $^4$He in Gelsil agrees semi-quantitatively with that in restricted geometries such as Vycor. 
 In our $^4$He-Gelsil system, however, the superfluid $T_{\mathrm c}$ is much lower than the peak temperature $T_{\mathrm B}$,
 so the nanoscale BEC occurs around $T_{\mathrm B}$ without macroscopic superfluid transition.
 
 Heat capacity peak without macroscopic superfluidity has been observed in liquid $^4$He droplets 
 formed in metal foils \cite{Syskakis}. In this case each droplet that is several nanometers in diameter
 is perfectly independent, and the droplets never exhibit superfluidity in macroscopic sense.
 The situation of $^4$He in nanoporous Gelsil is rather similar to this droplet system.
    
In the abovementioned LBEC scenario, the smallness of the pore size is only essential 
to the QPT behavior. 
It has been pointed out that disorder or randomness in porous structures produces Boson localization called Bose glass state\cite{Fisher}. 
In the Bose glass state, the condensates localize at the local minima of the random potentials, and macroscopic phase coherence 
is lost by the localization of atoms as in the case of narrowness-induced LBEC.
Kobayashi and Tsubota have recently studied superfluidity of $^4$He confined in a 3D random model potential taking account of 
the feature of our 2.5-nm Gelsil\cite{KobayashiTsubota}. 
They found that superfluidity disappears above 4.2 MPa due to the localization of the BECs.
It is in close agreement with our observation.

\subsection{Effects of Phase Fluctuations}
In the LBEC state, phase of the superfluid order parameter in each LBEC is fluctuating.
This phase fluctuation can contribute to the superfluid properties below $T_{\mathrm c}$.
We propose that the $T$-linear behavior in the super-(or normal) fluid density observed at high pressures 
(Fig. \ref{Coefficient}) is 
the manifestation of the phase fluctuations that are induced thermally (classically). 

The effects of phase fluctuation has been studied in Josephson junction arrays \cite{Mooij} 
and granular metal films\cite{Merchant}, which show a superconductor - insulator quantum phase transition 
by controlling some experimental parameters such as magnetic field.
It has also been proposed in the field of high-$T_{\mathrm c}$ cuprates that the phase fluctuations play an
important role on the properties of underdoped regimes. 
Emery and Kivelson (EK)\cite{Emery95} argued that low carrier-density superconductors such as high-$T_{\mathrm c}$ cuprates are characterized by a small phase stiffness,
and consequently the large phase fluctuations dominate notably the superconducting properties of underdoped regimes.
The emergence of the "pseudo-gap" states is caused by the local Cooper pairing without global phase coherence throughout the sample,
 and the onset of long range phase order determines the true superconducting transition that is detected by macroscopic means.
 This proposed mechanism is exactly the same as the LBEC picture in the $^4$He-nanopore system.
The LBEC state just corresponds to the EK pseudogap state.

The superfluid systems that are controlled by phase fluctuations possess the following characteristics\cite{Carlson99,Carlson05}:
\begin{enumerate}
  \item The superfluid density $\rho_{\mathrm s}$ is low, i.e. the phase stiffness (helicity modulus) is small even at 0 K. 
  \item The local order occurs at higher temperature than the long-range phase ordering.
\item If the phase fluctuation is thermally excited, $\rho_{\mathrm s}$ is proportional to $T$ at low $T$.
\item The long-range ordering $T_{\mathrm c}$ is proportional to $\rho_{\mathrm s}$. 
\end{enumerate}

In high-$T_{\mathrm c}$ cuprates, the $T$-linear behavior was observed by the measurement of penetration depth\cite{Hardy93,Roddick95}.
Although it is also attributed to the $d$-wave nature of the gap function, XY models with classical phase fluctuations
reproduce quite well the overall temperature dependence of $\rho_{\mathrm s}$\cite{Carlson99}.
The smallness of $\rho_{\mathrm s}$ and the proportionality between $\rho_{\mathrm s}$ and $T_{\mathrm c}$
was also confirmed and stressed as an important characteristic of various exotic superconductors by Uemura \textit{et al.}\cite{Uemura89,Uemura91}.
The proposed "universal" relation between $T_{\mathrm c}$ and muon relaxation rate was later reinterpreted by EK as an upper bound of 
$T_{\mathrm c}$ given by the phase-order temperature\cite{Emery95}.

All the abovementioned features of the phase-fluctuation model are actually observed in the $^4$He-Gelsil system we studied. 
As is shown in Figs. \ref{CRhoS} and \ref{Coefficient}, the $T$-linear behavior in $\rho_{\mathrm s}$ becomes prominent at pressures higher than 2.3 MPa.
This behavior strongly suggests the existence of classical phase fluctuations which dominates the normal fluid component.
It is also noted that the overall shape of the $\Delta f(T)$ curve bears striking resemblance 
to the superfluid density of measured in cuprates and the calculated one in the 3D XY model\cite{Carlson99}.

\begin{figure}[t]
 \centering
 \includegraphics[height=65mm]{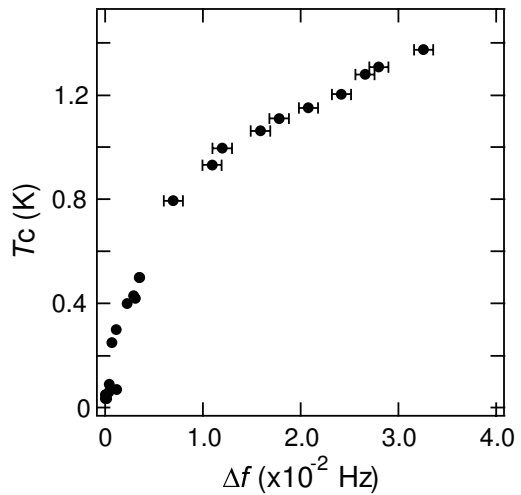}
  \caption{\label{Tc_Df}Superfluid transition temperature $T_{\mathrm c}$ of pressurized liquid as a function of the frequency shift $\Delta f$ at 10 mK.
  This plot corresponds to the Uemura plot for unconventional superconductors\cite{Uemura89,Uemura91}.}
\end{figure}

Moreover, a plot of $T_{\mathrm c}$ versus $\rho_{\mathrm s}$ (the so-called Uemura plot\cite{Uemura89,Uemura91}) 
in Fig. \ref{Tc_Df} clearly have tendencies that at $P > 2.3$ MPa $\rho_{\mathrm s}$ becomes small and approximately proportional to $T_{\mathrm c}$. 
The emergence of $T$-linear term in $\rho_{\mathrm s}$ correlates to the change in the slope of $T_{\mathrm c} - \rho_{\mathrm s}$ curve.

The accuracy in the determination of $T$-linear coefficient and $T_{\mathrm c} - \rho_{\mathrm s}$ relation
near $P_{\mathrm c}$ are degraded in our current torsional oscillator measurement because of the small $\rho_{\mathrm s}$
(small signal-to-noise ratio) and the coupling of oscillation to fourth sound. 
Measurements of $\rho_{\mathrm s}$ by other techniques such as fourth sound resonance method
will be essential.

The idea of LBEC gives a new perspective to a number of experimental studies of $^4$He 
in restricted geometries. 
The detailed torsional oscillator and specific heat studies by Reppy and coworkers\cite{Kiewiet75,Chan88,Zassenhaus99} 
shows that the superfluid transition occurs at slightly lower temperature than the 
temperature of the broad specific heat peak. 
At superfluid $T_{\mathrm c}$, an extremely small peak is additionally observed.
As well as in the Gelsil case, the broad peak is attributed to the formation of LBECs 
and the macroscopic superfluid transition occur at $T_{\mathrm c}$.
The LBEC picture in the $^4$He-Vycor system is also supported by 
the neutron and ultrasound experiment conducted by Glyde, Mulders and coworkers,
in which the roton signals are observed above $T_{\mathrm c}$ determined by ultrasound.
Thus, the "separation" of BEC and superfluid transition should be a universal characteristic 
of $^4$He in nanoporous media.

The $T$-linear superfluid density was observed in $^4$He filled in packed powders\cite{Pobell} and in 2.5-nm Gelsil\cite{Miyamoto96} at ambient pressures.
In these studies the $T$-linear behavior was attributed to the effect of "zero-dimensional" (0D) phonons.
However, the 0D phonons in the 3D connected nanopores are hard to imagine. It is rather reasonable
to interpret as the effect of phase fluctuations. Then a question arises:
Why is the $T$-linear behavior much more prominent in packed powder and MT's Gelsil than ours?
It is conjectured that difference in pore structure influence the Boson localization and thus the phase fluctuations.
Further studies including detailed characterization of porous materials are obviously intriguing. 
\section{Summary}
In summary, we have determined the anomalous phase diagram of $^4$He confined in the 2.5-nm Gelsil nanopores.
It is ultimately proven by torsional oscillator and heat capacity studies that BEC and superfluidity take place at separate temperatures.
Key physics to understand the phase diagram and superfluid properties is 
localization of Bose-Einstein condensates caused by confinement or disorder.
Striking similarity to the superfluid behavior in high-$T_{\mathrm c}$ cuprates may also be an 
important clue to elucidate the mechanism of quantum phase transition.
The entire phase diagram will be understood basically in terms of the phase fluctuation model 
that was proposed by Emery and Kivelson in the interpretation of phase diagrams of high-$T_{\mathrm c}$ cuprates.

Our study shows that $^4$He in nanoporous media is an illustrative example of strongly correlated Bosons 
in potential, which produce intriguing quantum phenomena.

This work is supported by the Grant-in-Aid for Priority Areas 
"Physics of Super-clean Materials", and Grant-in-Aid for 
Scientific Research (A) 
from MEXT, Japan.

\end{document}